\documentclass[onecolumn,aps,pra,superscriptaddress,showpacs,11pt,notitlepage]{revtex4-1}
\usepackage[latin9]{inputenc}
\usepackage{xcolor}
\usepackage{pdfcolmk}
\usepackage{mathtools}
\usepackage{amsmath}
\usepackage{amssymb}
\usepackage{graphicx}
\usepackage{esint}
\usepackage{soul}
\PassOptionsToPackage{normalem}{ulem}
\usepackage{ulem}
\usepackage{siunitx}

\usepackage[unicode=true,
 bookmarks=false,
 breaklinks=false,pdfborder={0 0 1},backref=false,colorlinks=false]
 {hyperref}
\makeatletter
\providecolor{lyxadded}{rgb}{0,0,1}
\providecolor{lyxdeleted}{rgb}{1,0,0}

\usepackage{amsfonts}
\usepackage{color}
\usepackage{xcolor}
\usepackage[mathscr]{euscript}%\usepackage[uft8]{inputenc}
\usepackage{rotating} %

 \newcommand{\ud}[1]{{#1^{\dagger}}}
\newcommand{\bra}[1]{\left\langle #1\right|}
\newcommand{\ket}[1]{\left| #1\right\rangle}

\newcommand{\Tr}{\mathrm{Tr}}

\makeatother

\begin{document}

\title{Entangling one polariton with a photon:\\
effect of interactions on a single-polariton quantum state
}

\author{Álvaro Cuevas}
\altaffiliation{Both authors contributed equally to this work}

\affiliation{CNR NANOTEC---Institute of Nanotechnology, Via Monteroni, 73100 Lecce, Italy}\affiliation{Dipartimento di Fisica, Sapienza University of Rome, Piazzale Aldo Moro, 2, 00185 Rome, Italy }

\author{Blanca Silva}
\altaffiliation{Both authors contributed equally to this work}
\affiliation{CNR NANOTEC---Institute of Nanotechnology, Via Monteroni, 73100 Lecce, Italy}\affiliation{Departamento de F\'{i}sica Teórica de la Materia Condensada, Universidad Autónoma de Madrid, 28049 Madrid, Spain}

\author{Juan~Camilo López~Carreño}

\affiliation{Departamento de F\'{i}sica Teórica de la Materia
  Condensada, Universidad Autónoma de Madrid, 28049 Madrid, Spain}
\affiliation{Faculty of Science and
  Engineering, University of Wolverhampton, Wulfruna St, Wolverhampton
  WV1 1LY, UK}

\author{Milena de Giorgi}
\email{milena.degiorgi@nanotec.it}
\affiliation{CNR NANOTEC---Institute of Nanotechnology, Via Monteroni, 73100 Lecce, Italy}

\author{Carlos~Sánchez~Muñoz}

\affiliation{Departamento de F\'{i}sica Teórica de la Materia Condensada, Universidad Autónoma de Madrid, 28049 Madrid, Spain}

\author{Antonio Fieramosca}

\affiliation{CNR NANOTEC---Institute of Nanotechnology, Via Monteroni, 73100 Lecce, Italy}

\author{Daniel G. Suárez-Forero}

\affiliation{CNR NANOTEC---Institute of Nanotechnology, Via Monteroni, 73100 Lecce, Italy}

\author{Filippo Cardano}

\affiliation{Università di Napoli Federico II, Napoli, Italy}

\author{Lorenzo~Marrucci}

\affiliation{Università di Napoli Federico II, Napoli, Italy}

\author{Vittorianna Tasco}

\affiliation{CNR NANOTEC---Institute of Nanotechnology, Via Monteroni, 73100 Lecce, Italy}

\author{Giorgio Biasiol}

\affiliation{Istituto Officina dei Materiali CNR, Laboratorio TASC, I-34149 Trieste, Italy}

\author{Elena del~Valle}

\affiliation{Departamento de F\'{i}sica Teórica de la Materia Condensada, Universidad Autónoma de Madrid, 28049 Madrid, Spain}

\author{Lorenzo Dominici}

\affiliation{CNR NANOTEC---Institute of Nanotechnology, Via Monteroni, 73100 Lecce, Italy}

\author{Dario Ballarini}

\affiliation{CNR NANOTEC---Institute of Nanotechnology, Via Monteroni, 73100 Lecce, Italy}

\author{Giuseppe Gigli}

\affiliation{CNR NANOTEC---Institute of Nanotechnology, Via Monteroni, 73100 Lecce, Italy}

\author{Paolo Mataloni}

\affiliation{Dipartimento di Fisica, Sapienza University of Rome, Piazzale Aldo Moro, 2, 00185 Rome, Italy }

\author{Fabrice~P.~Laussy}
\email{F.Laussy@wlv.ac.uk}
\affiliation{Faculty of Science and
  Engineering, University of Wolverhampton, Wulfruna St, Wolverhampton
  WV1 1LY, UK}
\affiliation{Russian Quantum Center, Novaya 100, 143025 Skolkovo,
  Moscow Region, Russia}

\author{Fabio Sciarrino}

\affiliation{Dipartimento di Fisica, Sapienza University of Rome, Piazzale Aldo Moro, 2, 00185 Rome, Italy }

\author{Daniele~Sanvitto}

\affiliation{CNR NANOTEC---Institute of Nanotechnology, Via Monteroni, 73100 Lecce, Italy}

\begin{abstract}
  % \textbf{Polaritons are quasi-particles originating from the
  % coupling of light with matter and that demonstrated quantum
  % phenomena at the
  % many-particle mesoscopic level, such as BEC and superfluidity.  A
  % highly sought and long-time missing feature of polaritons is a
  % genuine quantum manifestation of their dynamics at the
  % single-particle level.  Although they are conceptually perceived
  % as entangled states and theoretical proposals abound for an
  % explicit manifestation of their single-particle properties, so far
  % their behaviour remained fully accountable for by classical and
  % mean-field theories. In this Article, we report the first
  % experimental demonstration of a genuinely-quantum manifestation of
  % microcavity polaritons, by swapping the entanglement between a
  % polariton and an external photon from a photon pair generated by
  % parametric down-conversion.  Moreover we demonstrate how
  % interactions manifest at a single polariton level by perturbing
  % the entangled state with a rarefied polariton condensate. Our
  % results open the page of quantum polaritonics and implement the
  % first application of quantum spectroscopy.}
\end{abstract}

\date{\today}

\maketitle

\textbf{Polaritons are quasi-particles originating from the coupling
  of light with matter that demonstrated quantum phenomena at the
  many-particle mesoscopic level, such as BEC and superfluidity.  A
  highly sought and long-time missing feature of polaritons is a
  genuine quantum manifestation of their dynamics at the
  single-particle level.  Although they are conceptually perceived as
  entangled states and theoretical proposals abound for an explicit
  manifestation of their single-particle properties, so far their
  behaviour has remained fully accountable for by classical and
  mean-field theories. In this Article, we report the first
  experimental demonstration of a genuinely-quantum manifestation of
  microcavity polaritons, by swapping, in a two-photon entangled state
  generated by parametric down-conversion, a photon for a polariton.
  Furthermore, we show how single polaritons are affected by
  polariton-polariton interactions in a propaedeutic demonstration of
  their qualities for quantum information applications.}

Light would be the perfect platform for future quantum information
processing devices, would it not be for its too feeble
interaction.~\cite{obrien07a} A remedy is to rely on hybrid systems
that involve a matter component, bringing in strong
interactions.~\cite{cirac97b,reiserer14a} In the regime of strong
coupling that binds together light and matter, the resulting
polaritons appear as candidates of choices to deliver the strongly
interacting photons required in tomorrow's quantum
technology~\cite{reiserer15a} and indeed, a photon-photon gate relying
on these ideas has been recently demonstrated with a single atom in a
cavity.~\cite{hacker16a} At a theoretical level, the polariton is
itself an entangled superposition of light with a dipole-carrying
medium, of the form
$\ket{\mathrm{U}/\mathrm{L}}=\alpha\ket{0_{a}1_{b}}\pm\beta\ket{1_{a}0_{b}}$,
with $\ket{1_{a}}$ a photon and, depending on the system,
$\ket{1_{b}}$ a phonon, a plasmon, an exciton or even a full atom.
The exciton-polariton, that lives in
semiconductors,~\cite{kavokin_book11a} has already demonstrated to be
on par with cold-atoms physics, with reports that include
Bose-Einstein condensation~\cite{kasprzak06a} (up to room
temperature~\cite{christopoulos07a}),
superfluidity,~\cite{amo09a,amo09b} topological
physics~\cite{karzig15a} and an ever growing list of other exotic
quantum macroscopic phases.~\cite{carusotto13a} This 2D-particle, that
combines antagonist properties of light (lightweight and highly
coherent) and matter (heavy and strongly interacting) as it
propagates, also holds great promises for the future of classical
technology by allowing the shift from electronic to optical
devices~\cite{ballarini13a} and emerging as the future generation of
lasers.~\cite{schneider13a,kim16a} Thus, one of the major prospects of
polaritons is their quest for quantum-based technologies, as quantum
particles akin to photons but with much stronger nonlinearities.

Yet, even in configurations that are expected to provide strong
quantum correlations,~\cite{ardizzone12a} exciton-polaritons have so
far remained obstinately classically correlated.  However useful is
their description as a quantum superposition of excitons and photons,
with~$\alpha$, $\beta$ complex probability amplitudes, the actual
state typically realized in the laboratory is instead a thermal or
coherent distributions of such particles that yield, e.g., for the
case of coherent excitation, product
states~$\ket{\alpha}_{a}\ket{\beta}_{b}$.~\cite{dominici14a} The
variables~$\alpha$ and $\beta$ obey the same equations of motion and
evolve identically to quantum probability amplitudes, but they now
describe classical amplitudes of coherent fields for the photon~($a$)
and exciton~($b$), with no trace of purely quantum effect such as
entanglement and nonlocality.

This does not invalidate the interest and importance of reaching the
genuinely quantum-correlated regime of
polaritons~\cite{ciuti04a,savasta05a,demirchyan14a} (from now on, we
write polariton for exciton-polariton).  Following decades of
sustained efforts in this direction, the state-of-the-art is the
recently reported unambiguous squeezing of the polariton
field.~\cite{karr04a, boulier14a} This remains however a weak
demonstration of non-classicality (a classical field can be squeezed
in all its directions). What is instead required for a full
exploitation of quantum properties is a stronger class of quantum
states that are non convex mixtures of Gaussian states, i.e., akin to
Fock states and suitable to perform full quantum information
processing.~\cite{filip11a} Polariton blockade,~\cite{verger06a} that
relies on nonlinearities to produce antibunching, remained so far
unsuccessful.  Much hope was rekindled by the unconventional polariton
blockade~\cite{liew10a,bamba11a} based on destructive interferences of
the paths leading to two-excitation states, allowing much stronger
antibunching, although not in the sought class beyond convex mixtures
of Gaussian states.~\cite{lemonde14a} The race for quantum
polaritonics is thus still in its starting blocks and several
questions remain pending to justify polaritons as strong contenders
for quantum information processing.  Indeed, one of the main
uncertainties in the community is whether polaritons are robust
against pure dephasing. Being composed of interacting
particles---excitons---they could be strongly affected by the
environment in a harmful way for their quantum coherence.
Furthermore, even assuming that one could generate a polariton qubit,
another wishful quality of polaritons remains to be established: can a
single polariton be substantially affected by interactions with other
polaritons?  This is of paramount importance to use polaritons as
building blocks for nonlinear quantum gates.

\begin{figure}
\includegraphics[width=0.95\linewidth]{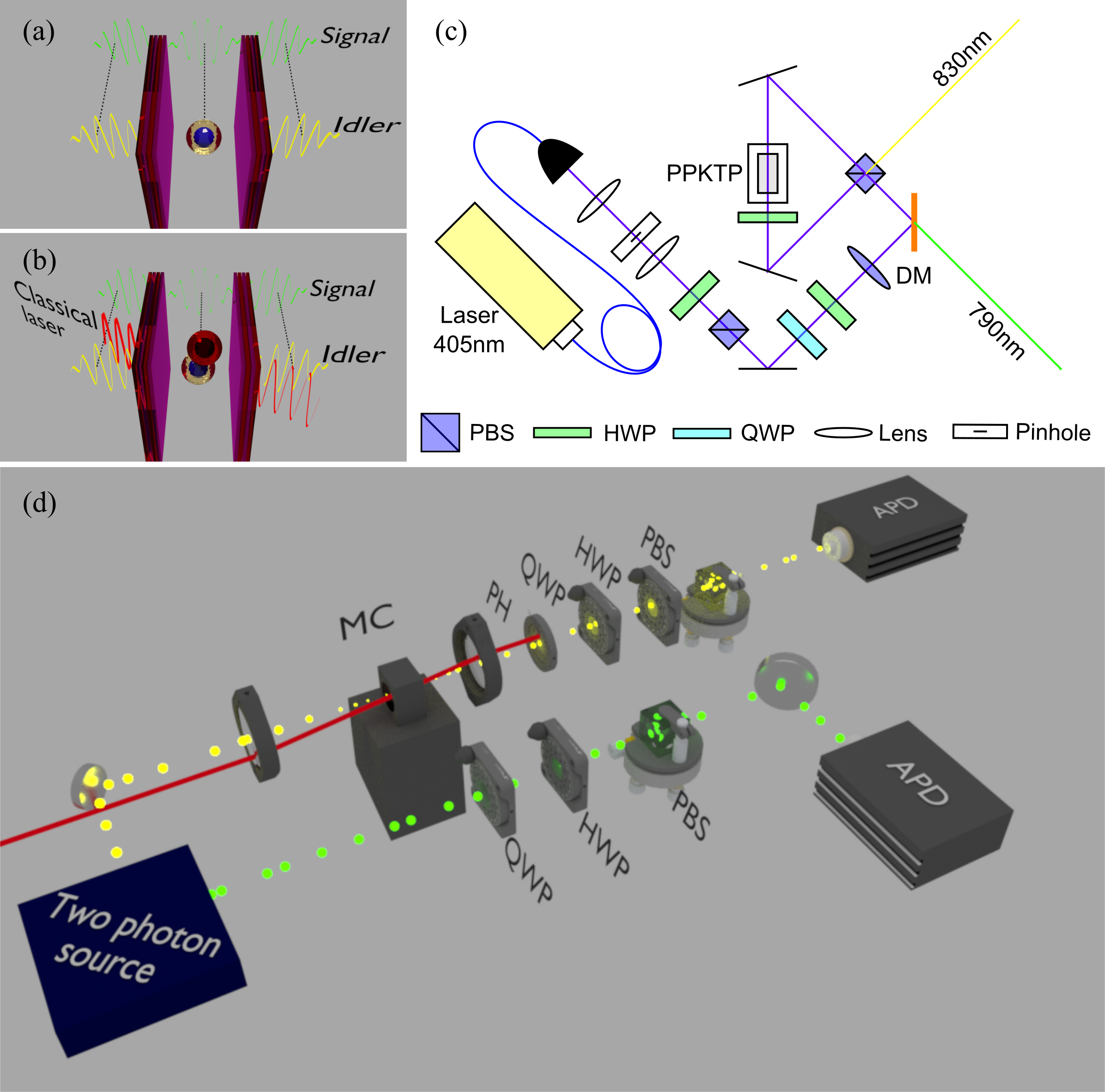}
\protect\caption{Sketch of the behaviour inside the microcavity: (a)
  Linear regime: a single photon (yellow, idler) gets into the
  microcavity where it becomes a polariton, that is later reconverted
  into an external photon. When this photon is entangled with another
  one (green, signal), we show that the entanglement is preserved
  throughout. (b) Interacting regime: a single photon (yellow, idler) enters
  the microcavity along with photons from a classical laser (red). The
  polaritons in the microcavity interact. This affects the
  entanglement of the single-polariton in a way that allows to measure
  the polariton-polariton interaction. (c) Sketch of the Sagnac
  interferometric source. (d) Sketch of the setup implementing these
  configurations.}
\label{fig:MCs} 
\end{figure}

In this Article, we answer positively to these two outstanding
questions, by turning to a new paradigm for creating quantum
polaritons. Instead of realizing a genuine quantum state from within,
e.g., based on strong nonlinearities, we imprint it from
outside.~\cite{lopezcarreno15a} Since photonics is so-far leading in
the generation, transfer and manipulation of quantum states, and since
photons couple well to polaritons, this opens an opportunity to
realize and explore quantum effects with polaritons, leaving to
technology to provide self-serviced samples in the future.  To ensure
the quantum nature of the polariton field, we use entangled photon
pairs: one photon is sent to the microcavity and the other is used to
later check that quantum correlations are still present and,
therefore, have been transferred to the polaritons.  The source of
entangled photon-pairs is a continuous wave (cw) laser at
$\lambda=\SI{405}{\nano\meter}$, down-converted from a Periodically
Poled KTP crystal (PPKTP)~\cite{fedrizzi07a} in order to generate
pairs of photons with a bandwidth narrow enough to couple to our
microcavity. The latter is described in the Methods.  The PPKTP is
introduced inside a Sagnac interferometer~\cite{fedrizzi07a} (see
sketch in Fig.~\ref{fig:MCs}) which allows to create polarization
entangled photons in a state of the
form~$\ket{\Psi}=(1/\sqrt{2})(\ket{\mathrm{H}\mathrm{V}}+e^{i\phi}\ket{\mathrm{V}\mathrm{H}})$,
where~$\ket{\mathrm{H}}$ stands for a horizontally polarized photon
and $\ket{\mathrm{V}}$ for a vertically polarized one. The phase can
be controlled to create a Bell state
$\ket{\Psi^{\pm}}=1/\sqrt{2}(\ket{\mathrm{H}\mathrm{V}}\pm\ket{\mathrm{V}\mathrm{H}})$
by placing a liquid crystal in one arm of the interferometer, at the
end of which a pair of non-degenerate entangled photons is emitted in
different directions: the idler at the wavelength of the polariton
resonances ($\lambda\sim\SI{830}{\nano\meter}$), and the signal at
higher energies ($\lambda\sim\SI{790}{\nano\meter}$). The high-energy
photon is then directed to a standard polarization tomography stage,
consisting of a quarter waveplate (QWP) followed by a half waveplate
(HWP), a polarizing beamsplitter (PBS) and a single-photon
detector~(APD). The low-energy photon, instead, is directed towards
the microcavity where it excites a single polariton, is stored there
until its eventual re-emission, is retrieved and finally directed to a
second tomography stage, equal to the one for the first photon. The
coincidence counts from the two single-photon detectors are measured
in a \SI{4}{\nano\second} window with a home-made coincidence unit. By
making such measurements for all the combinations of polarizations on
each detector (that makes~$6\times6$ possibilities for the three
polarization bases: circular, vertical and diagonal),~\cite{james01a}
we are able to determine whether the photon that transformed into a
polariton and back retained the nonlocal quantum correlations of a
Bell state. If this is the case, this proves that the polariton state
that inherited and later passed this information was itself in a
genuine one-particle quantum state, and was furthermore entangled with
the external photon propagating on the other side of the
setup. Namely, we have created the state
$\ket{\Psi^\pm}=(1/\sqrt{2})(\ud{a_{\mathrm{H}}}\ud{p_{\mathrm{V}}}\pm\ud{a_{\mathrm{V}}}\ud{p_{\mathrm{H}}})\ket{0}$
where $p_{\mathrm{H}}$ and $p_{\mathrm{V}}$ are the boson annihilation
operators for the horizontally and vertically polarized polaritons and
$a_{\mathrm{H}}$ and $a_{\mathrm{V}}$ for the signal photon.  While
this could be well expected from a standard linear theory of light
propagating through resonant dielectrics, the failures of polaritons
to demonstrate other expected quantum correlations, such as in OPO
scattering, show that it is not at all obvious that polaritons are
proper carriers of quantum features. In particular, their composite
nature and large exciton component could lead to fast decoherence
(e.g., via phonon scattering).

\begin{figure}
\includegraphics[width=0.95\linewidth]{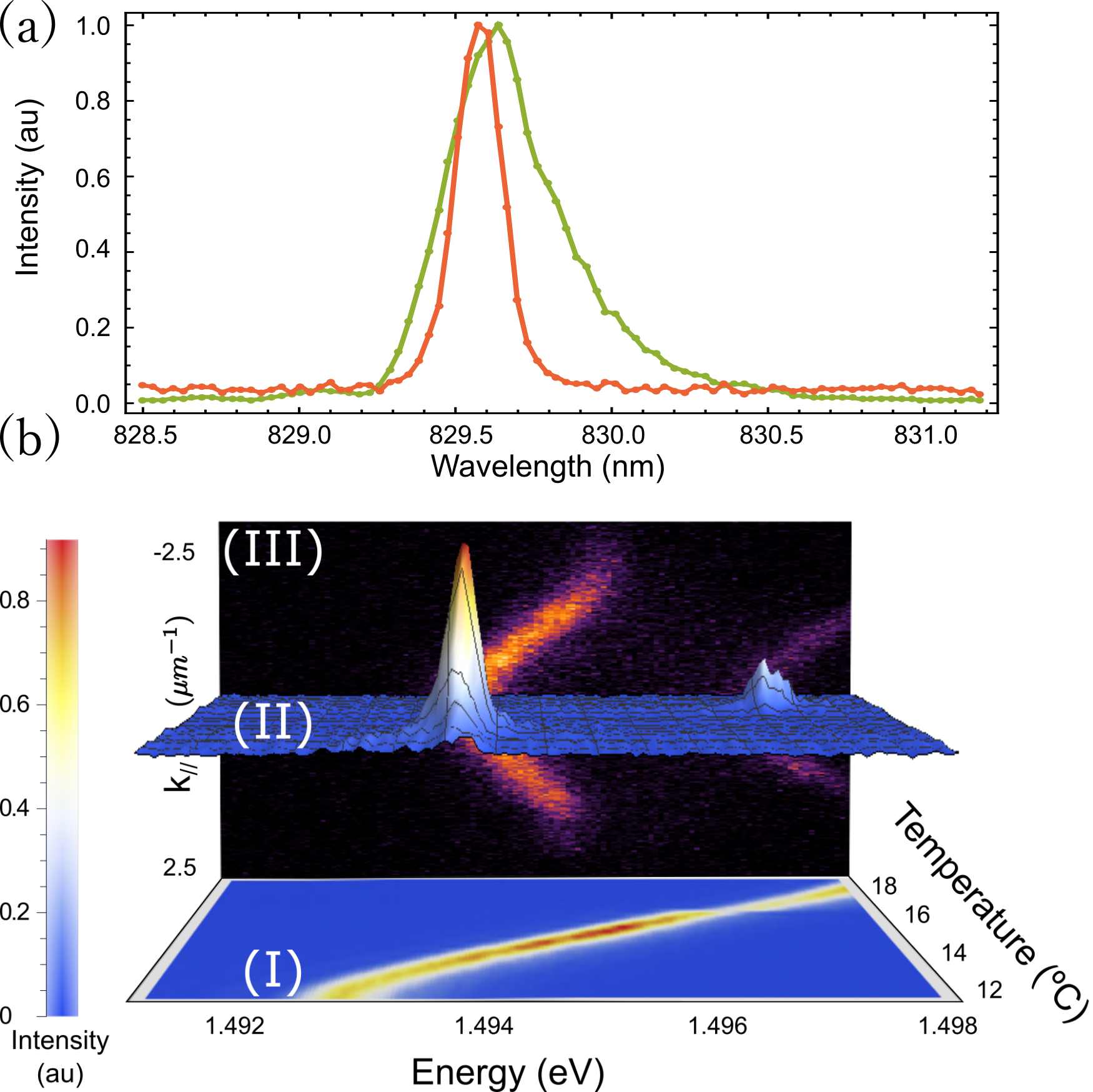}
\protect\caption{(a) Green: normalized emission of the PPKTP
  crystal. Orange: normalized transmission of the emission of the
  PPKTP crystal through the LPB. (b) Changes on the resonance as a
  function of the temperature of the two-photon source.  I: 2D map of
  the emission of the crystal outside of the microcavity. II:
  transmitted intensity as a function of the energy as the temperature
  varies. Two peaks can be identified that correspond to the
  resonances with one of the two polariton branches. The color-bar
  corresponds to both~I \&~II. III: far field of the emission under
  non-coherent pumping.}
\label{fig:resonances} 
\end{figure}

The spectral shape of the idler state is shown in
Fig.~\ref{fig:resonances}(a). Using single mode laser excitation of
the PPKTP crystal, the bandwidth of the entangled photon pairs is
reduced to \SI{0.46}{\nano\meter}, which is only 35\% wider than the
polariton state. Using temperature tuning on the nonlinear crystal we
could move the idler resonance from \SI{825}{\nano\meter} to
\SI{831}{\nano\meter}.  In order to check that the entangled idler is
transferred into the polariton state and not passing through the
cavity mode, we performed transmission measurements while scanning the
energy of the idler from below the lower polariton branch (LPB) to
above the upper polariton branch (UPB). In
Fig.~\ref{fig:resonances}(b), the effect of the microcavity on the
idler state shows that no light is transmitted when the idler is
out-of-resonance with the polaritons. This means that every photon
that passes through the sample has been converted into a polariton. If
this would not be the case, we would have observed at the cavity mode
(between the LPB and UPB) a finite signal, which is completely absent
even in logarithmic scale (not shown).

\begin{figure}
  \includegraphics[width=0.95\linewidth]{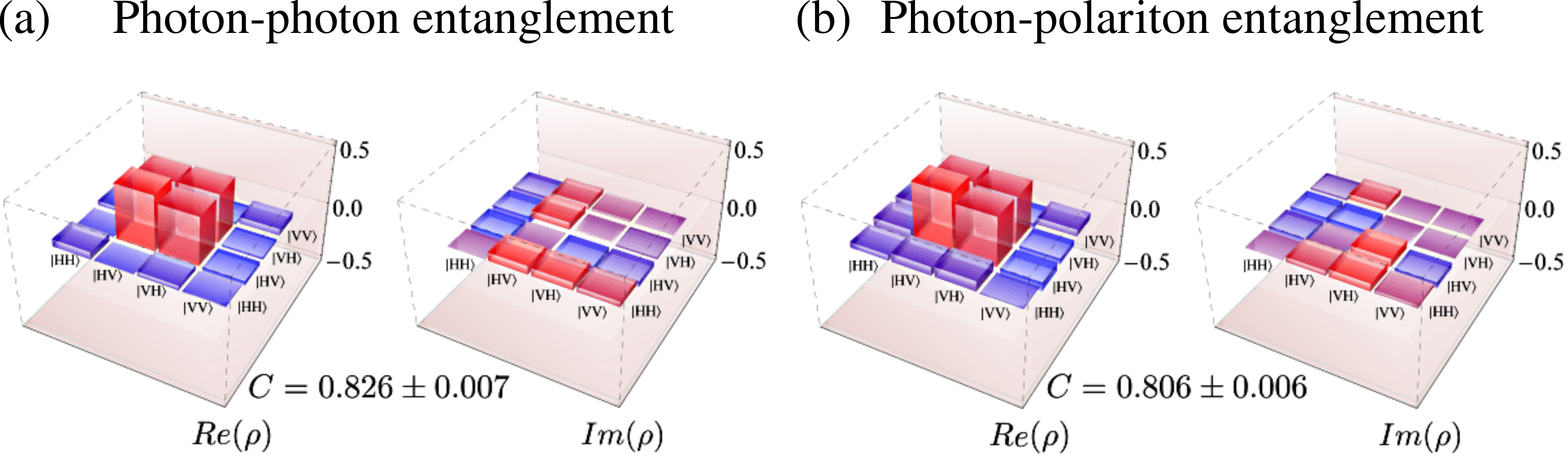}
\protect\caption{Tomography measured between the signal and idler
  photons. The signal is sent directly towards the detector while the
  idler photon becomes a polariton when entering the sample. (a) Real
  (left) and imaginary (right) components of the density matrix for
  the source of photon-pairs without the sample. The concurrence is
  not unity because the operation wavelength is not optimal for the
  source. (b) Real (left) and imaginary (right) components of the
  density matrix when passing through the microcavity. The concurrence
  of 0.806 shows that polaritons retain the entanglement.}
\label{fig:Alvaro} 
\end{figure}

For a bipartite system such as our photon pair, entanglement can be
quantified through the concurrence, defined in the Methods, which is
zero for classically correlated or uncorrelated states and gets closer
to one the higher the quantum entanglement.  To study the transfer of
entanglement into the polariton field, we compare the concurrence of
the two photons without the microcavity (corresponding to the
maximally entangled state) to that in which the photon is emitted by
the microcavity after the idler photon has been converted into a
polariton of the LPB with zero momentum ($k_{\parallel}=0$). As can be
seen in Fig.~\ref{fig:Alvaro}, the concurrence diminishes from 0.826
in the case of two freely-propagating photons, to 0.806 when one of
the photons is converted into a polariton. The maximum concurrence of
the PPKTP source is limited to 0.826 because the entangled state could
not be optimized at the operation wavelengths needed to interface
photons with polaritons.  Nevertheless, the concurrence is large and
remains so in presence of a microcavity in the way.  This result is a
direct proof that we have generated a single polariton within the
microcavity and that it existed there as a quantum state with no
classical counterpart.  If the single photon was lost in the polariton
thermal noise or coupled to a collection of other polariton states, no
entanglement would be eventually observed. This transfer of the photon
to and from a polariton state has shown to conserve almost entirely
the original degree of entanglement. Such a result is encouraging for
a future exploitation of quantum polaritonics.

\begin{figure}
\includegraphics[width=0.95\linewidth]{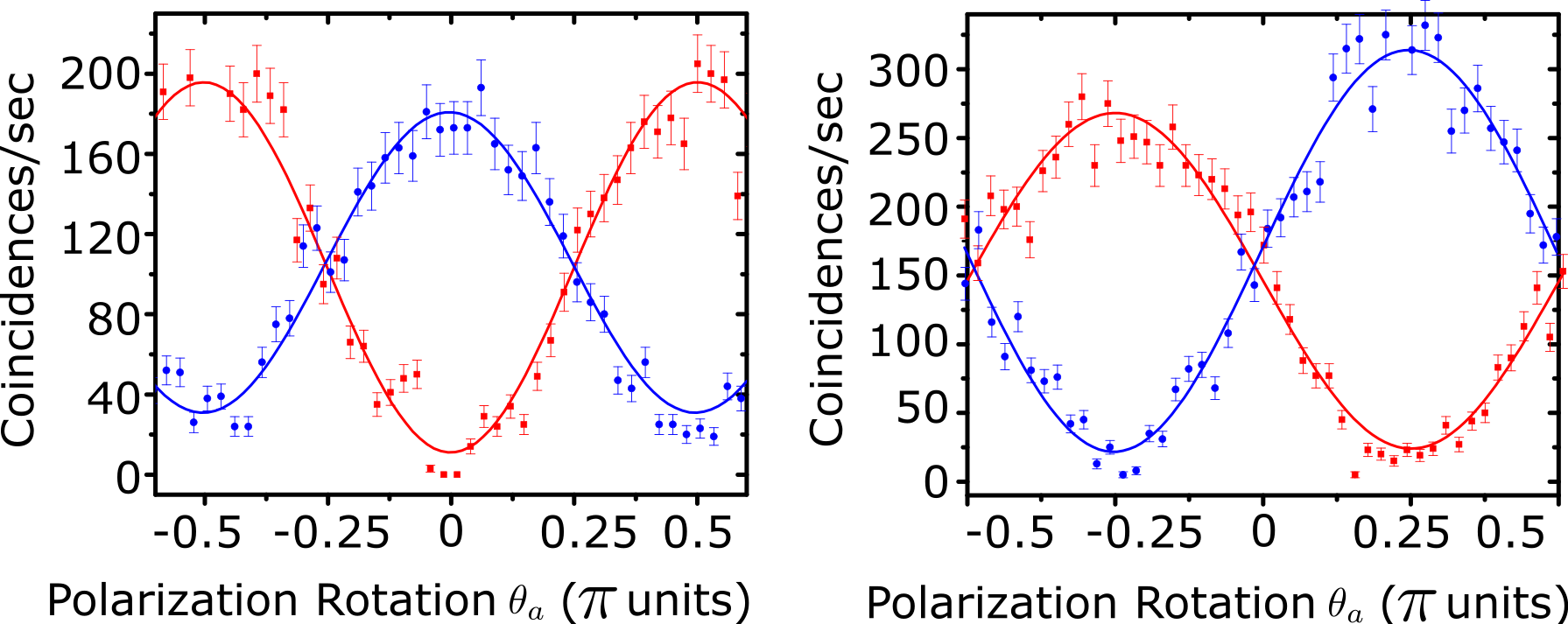}
\protect\caption{Coincidences as a function of the polarization
  between the external photons and the polaritons. Left: Bell curves
  for a polarization angle $b=0$. Right: Bell curves for
  $b=-\pi/4$. Red squares and blue circles denote the $++$ and $+-$
  coincidences, respectively.}
\label{Bell} 
\end{figure}

Beyond the preservation of concurrence, we also demonstrated the
nonlocality between the signal photon and the polariton created by
the idler photon. We did so by probing the classical
Clauser--Horne--Shimony--Holt (CHSH) inequality~\cite{clauser69a}
$S\le2$ where:
\begin{equation}
  S=|E(a,b)-E(a,b')+E(a',b)+E(a',b')|\,,
\end{equation}
with
\begin{equation}
  E(x,y)=
  \frac{C_{++}(x,y)+C_{--}(x,y)-C_{+-}(x,y)-C_{-+}(x,y)}{C_{++}(x,y)+C_{--}(x,y)+C_{+-}(x,y)+C_{-+}(x,y)}\,,
\end{equation}
a function of the coincident counts~$C_{pq}(x,y)$ between the two
measurement ports of our tomography setup, for the photons with
polarization~$p,q\in\{-,+\}$ in the bases~$x=a,a'$ and $y=b,b'$ where
$a=-\frac{\pi}{8}$, $a'=\frac{\pi}{8}$, $b=0$ and $b'=\frac{\pi}{4}$
are the combinations of polarization angles that maximize the Bell's
inequality violation. This can be obtained by rotating the tomography
HWPs in $a/2$, $a'/2$, $b/2$ and $b'/2$.  Figure~\ref{Bell} displays
the measured Bell curves, which show the photon coincidences
associated to these bipartite polarization measurements, along with
the continuous correlated oscillations predicted by the theory.  In
the optimum configuration, we obtain a value of $S = 2.463\pm 0.007$
which unambiguously violates the CHSH inequality and proves the
nonlocal character of the photon-polariton system.

The above experimental results clearly demonstrate that, although
polaritons are quasi-particles in a solid-state system with complex
and yet-to-be-fully-characterized interactions with their matrix and
other polaritons,~\cite{deveaud16a} they can be used as quantum bits
maintaining almost unvaried their quantum state and can transfer it
back and forth to an external photon. In particular, this shows that
several effects, such as pure dephasing, coupling to phonons,
radiative lifetime, etc., are not detrimental to quantum coherence.

To answer the second important question on the possibility to affect
the quantum state that we have created, we need to go into the
nonlinear regime and study the effect of polariton interactions.  To
observe such nonlinearities within our available equipment, we repeat
the experiment in a non polariton-vacuum configuration. Namely,
instead of exciting the sample with one photon of the PPKTP source
only, we add the excitation of a classical laser, which is the most
common way to excite a microcavity (see sketch in
Fig.~\ref{fig:MCs}). In this case, the entangled photon is sent on
resonance now with the UPB, and the classical laser with the LPB. We
have chosen this configuration to avoid any effect of relaxation of
the classical source into lower energy states while at the same time
keeping the quantum state well distinguishable from the other
polaritons. 

The classical pumping power is changed from having only the vacuum
state up to an average of 230 polaritons at any given time, with a
density still below $\SI{1}{polariton\per\square{\micro\meter}}$. The
calibration of the population is explained in the Methods.

In the conditions of our experiment, the full interacting-polariton
Hamiltonian can be reduced to the simple form (See Supplementary
Material):
\begin{equation}
  \label{eq:SatDec17170247CET2016}
  H= \omega_\uparrow \ud{q_\uparrow}q_\uparrow + \omega_\downarrow
     \ud{q_\downarrow}q_\downarrow + g_{\uparrow \downarrow}
     (\ud{q_\uparrow}q_\downarrow + \ud{q_\downarrow}q_\uparrow)\,,
\end{equation}
where~$q_p$ are quantized operators for the upper polariton at~$k=0$
with polarization $p=\uparrow$, $\downarrow$ and the lower polariton
condensate has been absorbed in the coefficients through a mean-field
approximation for the coherent
state~$\ket{\alpha_{\uparrow/\downarrow}}$ with
polarization~$\uparrow/\downarrow$:
\begin{subequations}
  \label{eq:viedic16211026CET2016}
  \begin{align}
   \omega_{\uparrow/\downarrow} &=
    \tilde\chi \big(3|\alpha|^2_{\uparrow/\downarrow}
    V^{(1)}+2|\alpha|^2_{\downarrow/\uparrow} V^{(2)}\big) + \omega_0\,,\\
    g_{\uparrow \downarrow} &= 2\tilde\chi{|\alpha_\uparrow \alpha_\downarrow|}V^{(2)}\,,
\end{align}
\end{subequations}
where $\tilde\chi\approx0.2$ and~$\omega_0$ are some constants linked
to the Hopfield coefficients that arise due to the geometry of the
experiment (see Supplementary) and, more importantly, $V^{(1,2)}$
correspond to same~(1) or opposite~(2) spin polariton-polariton
interactions. The nonlinear crystal emits pair of
polarization-entangled photons of the form:
\begin{equation}
  \label{eq:SatDec17172602CET2016}
  \ket{\psi_0} = \frac{1}{\sqrt{2}} \big (\ket{\mathrm{H,V}} +
  \ket{\mathrm{V,H}} \big) \equiv \frac{1}{\sqrt{2}} \big (
  \ud{c_\mathrm{H}} \ud{q_\mathrm{V}} + \ud{c_\mathrm{V}}
  \ud{q_\mathrm{H}} \big ) \ket{0}\,,
\end{equation}
where~$c_p$ is the quantum operator for the signal photon that goes
straight to the detector. The~$q_p$ photon, on the other hand, evolves
according to the Hamiltonian~(\ref{eq:SatDec17170247CET2016}). Even
for the free propagation $q^\dagger_pq_p$, the possible asymmetry for
the~$p=\mathrm{H}$ and~V polarizations results in different
phase-shifts, which alter the wavefunction as a whole. The rightmost
term in Eq.~(\ref{eq:SatDec17170247CET2016}), on the other hand,
results in a change of the state of polarization. By analyzing the
quantum correlations between the signal and idler after passing
through the cavity, one can thus gain information on the microscopic
parameters~$\omega_{\uparrow/\downarrow}$ and
$g_{\uparrow \downarrow}$.

\begin{figure}
  \includegraphics[width=0.95\linewidth]{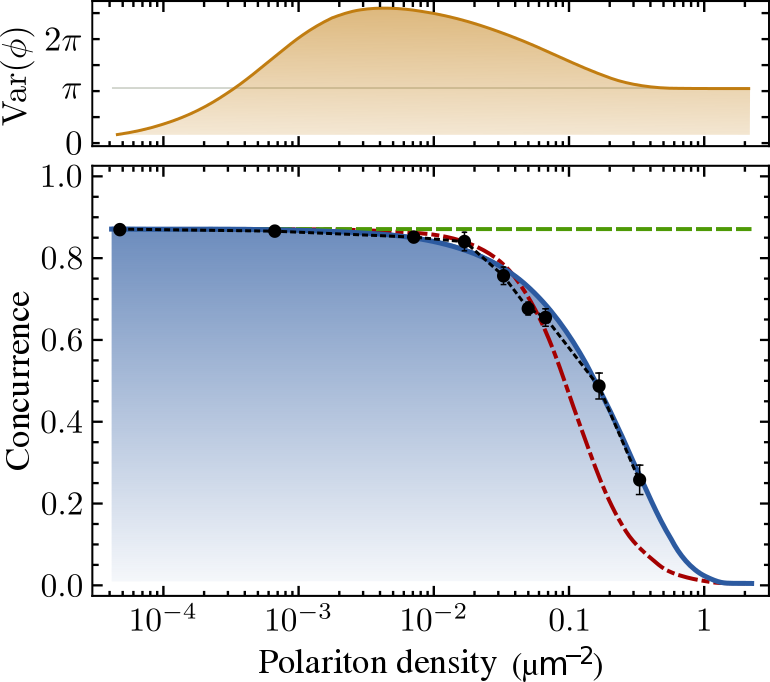}
  \protect\caption{Concurrence~$\mathcal{C}$ between the external
    photons and the polaritons as a function of the mean number of
    polaritons present in the sample (horizontal axis) from a
    classical laser. Each data point is obtained from the 36
    measurements of coincidences in all the combinations of
    polarization (dotted line serves as a guide). The solid line is a
    theoretical simulation for a model of fluctuating polarized lower
    polaritons that interact with the single upper polariton injected
    by the quantum source with an interaction strength of $16\%$~of
    the radiative broadening, and which is in excellent quantitative
    agreement with the observation. The orange line shows the
    variance of the phase-shift due to the fluctuations of the lower
    polaritons. The variance reaches its maximum when phase-shifts
    start to be stronger than~$2\pi$, thus causing a small dephasing.
    In the limit of very large fluctuations, the variance of the phase
    is that of a uniform distribution between $0$~and~$2\pi$,
    namely~$\sigma^2=\pi^2/3$ (thin horizontal gray line).  The green,
    dashed line is the theoretical model without the fluctuations,
    whereby the concurrence remains unaffected by the polariton
    condensate. The red, dashed-dotted line shows the theoretical
    simulation for a model of a condensate assuming thermal
    fluctuations, which fails to fit quantitatively the experimental
    observation and confirm that the condensate has Poissonian
    fluctuations, as expected.}
\label{fig:fig1} 
\end{figure}

At the level of the concurrence, our observation is unambiguous. As
shown in Fig.~\ref{fig:fig1} (data points), the concurrence decays
with increasing polariton density (following increasing laser
pumping). While the evolution of the wavefunction is expected, the
loss of concurrence seems to suggest a decoherence of the single
polariton when affected by interactions with the condensate. However,
this is due to a different scenario, beyond mean-field effects of the
lower polariton condensate, which, not being a Fock state, has
fluctuations (Poissonian ones)~\cite{love08a}. While in absence of
fluctuations, the model predicts that the state of the entangled
polariton acquires a relative phase-shift between the spin states of
the superposition (See Supplementary Material), which causes no loss
of concurrence (dashed green line), the averaging over several pairs
with a fluctuating phase-shift, concomitant with the fluctuating
condensate population, has the effect of an effective decoherence. In
this case, each entangled pair sees the microcavity in an
instantaneous different state, evolving the wavefunction differently
and acquiring a different relative phase-shift, which results in
spoiling the entanglement \textit{on average} as measured through the
signal/idler correlations in polarizations.  Implementing this effect
in the model allows us to reproduce theoretically the experimental
findings with an excellent agreement (solid blue line). We also show
the spread of the relative phase (orange dashed line) as measured by
its variance, that confirms that concurrence is lost due to a
fluctuating phase-shift, since the concurrence starts to decay when
the variance has reached its maximum, indicating that the phase is
randomly distributed (the variance decays because the phase-shift is
bounded modulo~$2\pi$; when fluctuations are so strong as to make the
phase-shift uniformly distributed, the variance becomes $\pi^2/3$ and
the concurrence vanishes).  This shows that the wavefunction of a
single-polariton Fock state is strongly affected indeed by its
interactions with the condensate in the LPB, albeit in a random way as
ruled by fluctuations currently beyond our control. However, in future
experiments, if the condensate would be replaced by another
non-fluctuating quantum state, such as another Fock state, this would
lead, instead of an apparent decoherence, to a well-controlled
spin-dependent phase-shift that would be able to power quantum
circuits. In fact, if the lower polariton branch would be in a Fock
state~$\ket{n_\uparrow\,,\,n_\downarrow}$
with~$n_{\uparrow,\downarrow}$ polaritons with polarization
$\uparrow,\downarrow$, respectively, the model shows, neglecting the
smaller~$V^{(2)}$ (see Supplementary Material) that a single polariton
in the upper polariton branch would then acquire a relative
phase-shift of:
\begin{equation}
  \label{eq:WedFeb15123032CET2017}
  \phi = 3 V^{(1)}\tau\bar\chi(n_\downarrow - n_\uparrow)\,,
\end{equation}
which, thanks to the large exciton-exciton repulsion~$V^{(1)}$,
results in a sizable shift, namely, with circularly-polarized
lower-polaritons only and for the parameters fitting the experiment,
$\phi\approx\frac{\pi}{30}n_\downarrow$.  With polaritons of
\SI{60}{\pico\second} lifetime, which is well within the state-of-the
art, one would thus get a deterministic $\pi$-shift for the single
polariton in the upper branch induced by one polariton only in the
lower branch, which would allow the realization of a polariton
controlled-Z gate, the fundamental building block for a controlled-NOT
(CNOT) gate. Our results are thus promising for on-chip implementation
of nonlinear quantum gates.

\begin{figure}
\includegraphics[width=0.95\linewidth]{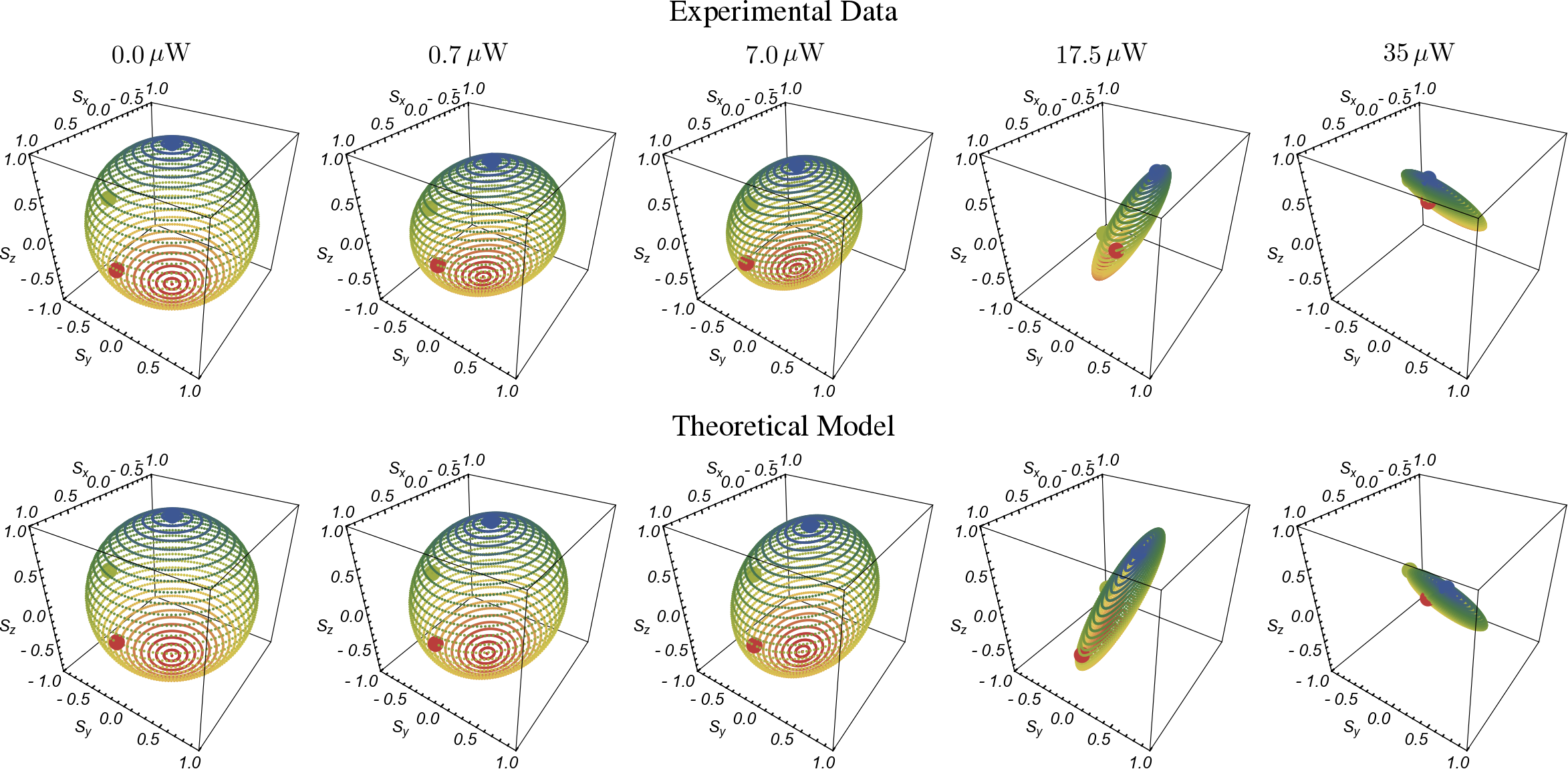}
\protect\caption{Ancilla assisted quantum process tomography. Our
  technique allows us to observe the effect of the lower polariton
  condensaste on the polarization of an upper polariton qubit, as a
  function of the increasing pumping (upper row). The shrinking of the
  Poincar\'e sphere into a spindle confirms that the qubit is affected
  by its interaction with the lower condensate, whose fluctuations
  lead to an effective decoherence when averaging. The fact that the
  decoherence occurs along a main axis $x$--$y$ suggests that the
  condensate experiences a renormalization of its state in this
  particular basis. Making this assumption in the theoretical model
  leads to an excellent agreement of the observed phenomenology (lower
  row) and points at future methods for characterizing also the
  condensate itself. The figures for the theoretical model were
  obtained for $0$, $0.018$, $0.032$, $0.226$ and~$0.272$ polaritons
  per micrometer squared, respectively.}
\label{fig:spheres} 
\end{figure}

One can go even further in characterizing the underlying Hamiltonian
by mapping its effect on all the possible polarized qubit states,
i.e., how it transforms the Poincar\'e sphere. When in possession of
entangled states, as in our case, this can be achieved with a single
input state only, thanks to a technique (further detailed in the
Methods) known as ancilla assisted quantum process
tomography.~\cite{altepeter03a} In essence, the sphere rotates under
unitary transformation and shrinks under the effect of decoherence.
Our joint experimental/theoretical analysis, presented in
Fig.~\ref{fig:spheres}, shows that the sphere does not rotate at lower
pumping, although it experiences some wobbling, until, at larger
pumping, it shrinks into a spindle, whose main axis does rotate.  The
experimental tomography is a direct post-processing of the data. The
theory applies the same procedure to input
states~(\ref{eq:SatDec17172602CET2016}) undergoing the evolution of
Hamiltonian~(\ref{eq:SatDec17170247CET2016}). We find that the
shrinking into a spindle, also due to the fluctuating condensate that
causes the loss of concurrence, can be obtained with no rotation of
the sphere with increasing pumping
when~$\omega_\uparrow-\omega_\downarrow$ and~$g_{\uparrow\downarrow}$
remain constant on average. In these conditions, both polarization
states at the extremities of the spindle have a constant energy shift
and there is no admixture of polarization. This results in decohering
all polarization states except in the $\uparrow$, $\downarrow$ basis,
while not producing a unitary rotation. The fact that the spindle
shrinks along an axis that varies with power suggests that at higher
pumping, the polariton condensate renormalizes itself into a
condensate of polarized quasi-particles in an arbitrary $x$, $y$
polarization basis, so that the effective Hamiltonian for the upper
polariton polariton becomes
$H =\omega_\mathrm{x}\ud{q_\mathrm{x}}q_\mathrm{x} + \omega_\mathrm{y}
\ud{q_\mathrm{y}} q_\mathrm{y} + g_\mathrm{xy} \ud{q_\mathrm{x}}
q_\mathrm{y} + g_\mathrm{xy}^\ast \ud{q_\mathrm{y}} q_\mathrm{x}$.
This introduces new microscopic parameters, $\omega_{x/y}$
and~$g_{xy}$, which, when kept constant in average with increasing
pumping but still fluctuating, reproduce the experimental finding
(cf.~Fig.~\ref{fig:spheres}). The details of the microscopic origin of
these terms are given in the Supplementary material. They are a
complex admixture of the interaction strengths and the state of the
condensate which in principle, could also be studied for their own
sake, although this is out of reach of the present work.  This shows,
however, that the technique also opens new perspectives for probing
classical features of the polaritons, such as the nature of the
polariton condensate and of its interactions, known to depart from the
atomic paradigm.~\cite{dominici15a}

In conclusions, the observation of a genuinely quantum state of the
polariton field has been demonstrated. Our experiments also bring to
semiconductors an implementation of quantum
spectroscopy~\cite{kira06a,assmann11b,mukamel15a,lopezcarreno15a} by
using quantum-correlated light to access phenomena and information out
of reach of a classical laser.  Our measurement in the presence of a
polariton condensate is the first of its kind and already confirms
that polaritons are serious players on the quantum scene through their
ability to exhibit strong nonlinear effects involving a
single-polariton state.  This has obvious implications for the design
and implementation of a new generation of quantum gates, routing
strongly-interacting polaritons in pre-determined landscapes so as to
make them interfere~\cite{sturm14a}. Our experiment might well be the
precursor for the long-sought strongly interacting photons needed for
the realization of scalable and efficient quantum computers.

\section*{Methods}

\subsection{Theoretical model}

The theory models the experiment by feeding
Hamiltonian~(\ref{eq:SatDec17170247CET2016}) (see Supplementary for
its derivation from the full polariton Hamiltonian) with the initial
state~(\ref{eq:SatDec17172602CET2016}) and acquiring statistics over
repetitions of this scenario with
coefficients~(\ref{eq:viedic16211026CET2016}) fluctuating with
Poissonian distributions with a
mean~$|\alpha_{\uparrow\downarrow}|^2$, modeling a polariton
condensate beyond mean-field.
 
The entanglement for bipartite systems can be quantified unambiguously
with the ``concurrence'', which extracts the amount of non-classical
correlations from an averaged density matrix:
\begin{equation}
\label{eq:juenov3155511CET2016}
\bar{\rho}\equiv \frac{1}{\mathcal{N}} 
\begin{pmatrix} \mathrm{HHHH} & \mathrm{HHVH} & \mathrm{HHHV} & \mathrm{HHVV} \\ h.c. & \mathrm{HVVH} & \mathrm{HVHV} & \mathrm{HVVV} \\ h.c. & h.c. & \mathrm{VHHV} & \mathrm{VHVV} \\ h.c. & h.c. & h.c. & \mathrm{VVVV}
 \end{pmatrix}\, ,
\end{equation}
where~$\mathcal{N}$ is a constant put here so that $\Tr(\bar\rho)=1$
and, e.g., VHHV stands for:
\begin{equation}
  \label{eq:mardic20170159CET2016}
   \frac{1}{N}\sum_{j=1}^{N}\bra{\psi_j}  \ud{c_\mathrm{V}} \ud{q_\mathrm{H}}
q_\mathrm{H} c_\mathrm{V} \ket{\psi_j}\,,
\end{equation}
where the states~$\ket{\psi_j}$ are those computed from the
fluctuating Hamiltonian as explained in the Supplementary
Material. This corresponds to the experimental configuration where
every element of the density matrix is reconstructed through a
tomographic process that requires up to~36 measurements detecting
every possible combination of polarization. From the density matrix in
Eq.~(\ref{eq:juenov3155511CET2016}) we compute the concurrence
as~$\mathcal{C}[\bar{\rho}]
\equiv\max(0,\lambda_1-\lambda_2-\lambda_3-\lambda_4)$ where
the~$\lambda_i$ are the eigenvalues in decreasing order of the
matrix~$\sqrt{\sqrt{\bar{\rho}}\tilde\rho\sqrt{\bar\rho}}$.
Here~$\tilde\rho\equiv(\sigma_y\otimes\sigma_y)
\bar{\rho}^T(\sigma_y\otimes\sigma_y)$, and $\sigma_y$ is a Pauli spin
matrix.

The \emph{Ancilla Assisted Quantum Process Tomography} (AAQPT)
characterises the action of an unknown map (black box) acting on one
of two entangled qubits, by using only one input and one output state
to recreate the associated map. This is thanks to the intrinsic
correlations of the bipartite state that dispense from considering the
transformation of a large collection of input states.

Since any qubit~$\ket{\psi}$ encoded in the polarization degree of
freedom has a Stokes decomposition of its associated density matrix
$\rho = \frac{1}{2}\vec{r}\cdot\vec{\sigma} =\frac{1}{2} (r_0
\mathbf{1} + r_1 \sigma_1 + r_2 \sigma_2 + r_3 \sigma_3)$ where
$\vec{r}$~is the Stokes vector and $\vec{\sigma}$~is the vector of
Pauli matrices,~\cite{james01a} AAQPT allows to obtain the black box
(the microcavity with a lower polariton condensate in our case)
map~$\chi$ from the expression
$\vec{r}_\mathrm{out} = \chi \vec{r}_\mathrm{in}$ where
$\vec{r}_\mathrm{in}$ and~$\vec{r}_\mathrm{out}$~are the Stokes
parameterization of the state of the single photon before and after it
goes through the cavity, respectively. The numerical calculation
of~$\chi$ is obtained by the linear
decomposition~$\chi=(A^{-1}B)^\mathrm{T}$ with~$A$ and~$B$ obtained
from the state of the pair of entangled photons before and after one
of the photons has gone though the cavity and are defined as
$A_{i,j} =\Tr [(\sigma_i \otimes \sigma_j)\rho_\mathrm{out}]$ and
$B_{i,j} =\Tr [(\sigma_i \otimes \sigma_j)\rho_\mathrm{in}]$.  In the
experiment, where errors and noise could lead to a non-physical
result, the map is best estimated by a Maximum Likelihood process
over~$\chi$. This technique minimizes the statistical dispersion of
the expectation values between~$\chi_\mathrm{raw}$ and its trace
preserving and completely positive version~$\chi_\mathrm{physical}$.
An intuitive representation of the map~$\chi$ that facilitate its
interpretation is given through the Poincar\'e sphere, where any qubit
is mapped as a point on the surface, using~$(r_1, r_2, r_3)$ as
$(x, y, z)$ coordinates (Fig.~\ref{fig:spheres}). There, $r_0$
represents the intensity of the field, $r_1$ the degree of linear
polarization, $r_2$ that of diagonal polarization and~$r_3$ that of
circular polarization.

\subsection{Sample \& Setup}

The microcavity sample is composed of front and back Distributed Bragg
Reflectors (DBR) with 20 pairs each, confining light, and one
In$_{0.05}$Ga$_{0.95}$As quantum well (QW), confining excitons. The QW
is placed at the antinode of the cavity to maximize their interaction
and enter the strong-coupling regime.~\cite{weisbuch92a} The
experiment consist in four different parts: photon generation,
signal-photon tomography stage, polariton source and polariton
tomography stage.  The first part consists of a Sagnac interferometer
excited with a single-mode diode laser at \SI{405}{\nano\meter} and a
pumping power of \SI{6.5}{\milli\watt} and bandwidth FWHM $<$
\SI{5}{\pico\meter}. Its power selection is achieved by fixing a
half-wave plate (HWP) before a polarizing beam splitter (PBS), and the
horizontal output polarization is again rotated in a desired arbitrary
polarization before a dichronic mirror (DM). In order to increase the
efficiency of the PPKTP, a lens is introduced to focalize the diode
right on the crystal. The second and the fourth sections of the setup
are used for the tomography measurement.  The idler photon goes
through the polariton source, which consists on a cryostat at
\SI{20}{\kelvin} and a pressure of \SI{100}{\milli\bar}. The
single-photon is focused on the surface of the microcavity and the
emitted photons are recollected with a second lens sending them
directly to the fourth stage (the idler's tomography stage). In the
experiment measuring nonlinear effects, an additional laser is sent to
the microcavity at a particular angle. The transmitted photons given
by the laser are covered with a diaphragm (pin-hole, PH) at the
Fourier plane of the recollection lens. The entanglement measurement
takes place in the tomography analysis.~\cite{altepeter05a} In our
case, we apply a hypercomplete tomography by projecting the bipartite
state onto a combination set of three bases: logical
($|\mathrm{HVHV}\rangle$ \& $|\mathrm{V}\rangle$), diagonal
($|+\rangle=|\mathrm{HVHV}\rangle+|\mathrm{V}\rangle$ \&
$|-\rangle=|\mathrm{HVHV}\rangle-|\mathrm{V}\rangle$), and circular
($|R\rangle=|\mathrm{HVHV}\rangle+i|\mathrm{V}\rangle$ \&
$|L\rangle=|\mathrm{HVHV}\rangle-i|\mathrm{V}\rangle$). Each local
projection is done by applying a rotation in a QWP and a HWP, followed
by a PBS as polarization filter (see Fig \ref{fig:MCs}). Finally, the
remaining photons belonging to the qubit of both the signal and the
idler are coupled to a single-mode optical fiber (SMF), connected to
an avalanche photo-detector (APD). The reconstruction of the state
sees the relative coincident count events (CC) among all the mentioned
projections during the desired integration time. In that way, we
measure the state by projecting many copies of the same.  The
measurements corresponding to the results reported in
Fig. \ref{fig:fig1} were unavoidably affected by the presence of the
classical cw laser which was minimised by momentum selection of the
polariton signal (as shown in Fig. \ref{fig:MCs}d). The contribution
of this noise, however, was subtracted from the raw data by performing
desynchronised tomography for each power of the external laser.  The
calibration of the population and density in presence of the classical
laser was obtained by means of four different parameters: the pumping
power, the photon energy, the polariton lifetime and the laser spot
size. The amount of photons/sec delivered by the laser to the
microcavity were calculated as the ratio Power/Energy. Given the
transmission of the microcavity and assuming the same amount of
photons to be emitted on both sides of the sample, only~0.5\% of the
photons become polaritons. Multiplying this rate of polaritons/sec by
the lifetime of a polariton (measured as \SI{2}{\pico\second}), gives
a maximum of 230~polaritons for the highest power used. The polariton
density polaritons was obtained by dividing by the spot
area~(\SI{706}{\micro\meter\squared}).

\section*{Acknowledgements}

This work was supported by the ERC-Starting Grants i) POLAFLOW (grant agreement no.~308136, http://polaritonics.nanotec.cnr.it) and ii) 3D-QUEST
(3D-Quantum Integrated Optical Simulation; grant agreement no.~307783, http://www.3dquest.eu).
It was also partially supported by PhD Chilean Scholarships CONICYT,
``Becas Chile'' and by the Spanish MINECO under contract
FIS2015-64951-R (CLAQUE).

\section*{Author contributions}
%All authors contributed equally to this work. 
DS and FPL conceived the idea. AC, BSF, MD, FC and DGS prepared the set up and performed the experiments with help from DB and LD. VT, GB grew the sample and AF prepared the sample for transmission measurements.  JCLC, CSM, EdV developed the theory under the supervision of FPL. LM, PM and FS supervised the experiments on photon-photon and photon-polariton concurrence. DS coordinated and supervised the project. All authors contributed to the analysis and interpretation of the data as well as the editing of the manuscript.

Authors declare no financial competing interest.

\newpage
%\section*{References}
\bibliographystyle{naturemag}
\bibliography{Sci,books,arXiv}

\end{document}